\documentclass[a4paper]{jpconf}
\usepackage{graphicx}
\usepackage{amssymb}

\begin{document}

\title{Transport coefficients and quark-hadron phase transition(s) from PLSM in vanishing and finite magnetic field}

\author{Abdel Nasser TAWFIK}

\address{Egyptian Center for Theoretical Physics (ECTP), Modern University for Technology and Information (MTI), 11571 Cairo, Egypt}
\address{World Laboratory for Cosmology And Particle Physics (WLCAPP), 11677 Cairo, Egypt}

\ead{a.tawfik@eng.mti.edu.eg}

\begin{abstract}

In peripheral heavy-ion collisions, localized ,short-lived an extremely huge magnetic field can be generated. Its possible influences on the quark-hadron phase transition(s) and the transport properties of the hadronic and partonic matter shall be analysed from the Polyakov linear-sigma model. Our calculations are compared with recent lattice QCD calculations.
 
\end{abstract}

\section{Introduction}

The heavy-ion collision can be divided into several successive processes including formation of quark-gluon plasma (QGP) and hadronization, etc. Some of such processes are believed to exist shortly after the big bang and explain the interior of compact stellar objects. On the other hand, the experimental devices are exclusively designed to detect hadrons, leptons and electromagnetically interacting particles. Thus, quarks and gluons can't directly be detected and therefore properties such as viscosity of QGP still represent a great challenge for the particle scientists. The nonnegligible elliptic flow measured at the Relativistic Heavy-Ion Collider (RHIC) and recently at the Large Hadron Collider (LHC) is a striking observation referring to the fact that the viscosity property turns to be measurable in the heavy-ion collision experiments, on one hand and the hydrodynamic response of produced matter to the initial geometry leads to such particle production asymmetry, on the other hand. 

Most of the experimental signals revealing the QGP formation and the elliptic flow, for instance, strengthen with increasing the collision centrality. Furthermore, the peripheral collisions introduce remarkable effects, as well. The local imbalance in the momentum that carried by the colliding nucleons generates a nonvanishing local angular-momentum \cite{prephiral2,prephiral3} and the relativistic motion of the spectator nucleons generates currents of net positive charge moving off-center (opposite to each other). Both types of nuclear collisions generate a huge magnetic field \cite{prephiral1}, ${\cal O}(m_{\pi}^2)$. 

In quantum electrodynamics (QED), phenomena such as magnetic catalysis in which the magnetic field dynamically generates masses, and Meissner effect where the magnetic field changes the order of the phase transition in type-I superconductor from second to first  are examples on the possible influence of such large magnetic field on the phase-transitions in quantum choromodynamics (QCD) and/or the properties of the hadronic and partonic matter. 

Influence of this magnetic field on the chiral phase-transition has been examined by using the linear-sigma model \cite{fraga}. In recent works, we have implemented the Polyakov linear-sigma model (PLSM) in characterizing various QCD phenomena and quantities. 
Thermodynamic quantities including normalized and non-normalized higher-order moments of the particle multiplicity have been investigated \cite{Tawfik:2014uka,Tawfik:quasi}. The chiral phase-structure of (pseudo)-scalar and (axial)-vector meson masses in thermal and dense medium has been determined \cite{Tawfik:Masses}. Also, it was reported on some properties of QGP in nonzero magnetic field \cite{Tawfik:2014hwa}. 

\section{Polyakov linear-sigma model in nonzero magnetic filed}

Let us assume that the generated magnetic field ($e B$) is directed along $z$-axis. From magnetic catalysis \cite{Shovkovy2013} and by using Landau quantization, the magnetic field modifies the quarks dispersion relations, 
\begin{eqnarray}
E_{B,f} &=& [p_{z}^{2}+m_{f}^{2}+|q_{f}|(2n+1-\sigma) B]^{1/2}, \label{eq:moddisp}
\end{eqnarray}
where $f$ runs over all quarks flavors, $n$ is the quantized Landau number, $\sigma$ is related to the spin quantum number and $S$; $\sigma=\pm S/2$.  
From dimensional reduction of the magnetic catalysis property \cite{Shovkovy2013}, 
\begin{eqnarray}
\int \frac{d^3 p}{(2 \pi)^3} &\longrightarrow& \frac{|q_{f}| B}{2 \pi} \sum_{\nu=0}^{\infty} \int \frac{d p_z}{2 \pi} (2-1 \delta_{0\nu}),
\end{eqnarray} 
where $2-1 \delta_{0\nu}$ stands for degenerate Landau levels and $2n+1-\sigma$ is replaced by the quantum number $\nu$

In mean field approximation, the PLSM grand potential is given as \cite{Tawfik:2014uka},
\begin{eqnarray}
\Omega(T, \mu) &=& U(\sigma_x, \sigma_y) + \mathbf{\mathcal{U}}(\phi, \phi^*, T) + \Omega_{\bar{\psi}
\psi} (T;\phi,\phi^{*},B).
\end{eqnarray} 
The mesonic potential $\mathbf{\mathcal{U}}(\phi, \phi^*, T)$ is assumed to vanish at high temperatures $\gtrapprox \Lambda_{QCD}$. The purely mesonic potential is given in Ref. \cite{Schaefer:2008hk}. Assuming Landau quantization and from magnetic catalysis, the quarks and antiquark contribution to PLSM potential becomes \cite{Tawfik:2014hwa} 
\begin{eqnarray}
\Omega_{\bar{q}q}(T, \mu _f, B)&=& - 2 \sum_{f=l, s} \frac{|q_f| B \, T}{(2 \pi)^2} \,  \sum_{\nu = 0}^{\nu _{max_{f}}}  (2-\delta _{0 \nu })    \int_0^{\infty} dp_z \nonumber \\ && \hspace*{5mm} 
\left\{ \ln \left[ 1+3\left(\phi+\phi^* e^{-\frac{E_{B, f} -\mu _f}{T}}\right)\; e^{-\frac{E_{B, f} -\mu _f}{T}} +e^{-3 \frac{E_{B, f} -\mu _f}{T}}\right] \right. \nonumber \\ 
&& \hspace*{3.7mm} \left.+\ln \left[ 1+3\left(\phi^*+\phi e^{-\frac{E_{B, f} +\mu _f}{T}}\right)\; e^{-\frac{E_{B, f} +\mu _f}{T}}+e^{-3 \frac{E_{B, f} +\mu _f}{T}}\right] \right\},\hspace*{5mm}  \label{new-qqpotio}
\end{eqnarray}
where $E_{B,f}$ is given in Eq. (\ref{eq:moddisp}) and $\mu_f$ refers to the chemical potential of $f$-quark.  In vanishing magnetic field, the quarks potential reads
\begin{eqnarray} 
\Omega_{\bar{q}q}(T, \mu _f)&=& -2 \,T \sum_{f=l, s} \int_0^{\infty} \frac{d^3\vec{p}}{(2 \pi)^3} \left\{ \ln \left[ 1+3\left(\phi+\phi^* e^{-\frac{E_f-\mu _f}{T}}\right)\times e^{-\frac{E_f-\mu _f}{T}}+e^{-3 \frac{E_f-\mu _f}{T}}\right] \right. \nonumber \\ 
& & \hspace*{34.mm} \left.  + \ln \left[ 1+3\left(\phi^*+\phi e^{-\frac{E_f+\mu _f}{T}}\right)\times e^{-\frac{E_f+\mu _f}{T}}+e^{-3 \frac{E_f+\mu _f}{T}}\right] \right\}, \hspace*{7mm}  \label{thermalOMG}
\end{eqnarray}

\section{Results and discussion}

From Green-Kubo formula, the bulk viscosity can be related to the correlation functions of trace of the energy-momentum tensor.
\begin{eqnarray}
\xi &=& \lim_{\omega \rightarrow\, 0}\frac{1}{9\, \omega} \int_0^\infty d t \, \int d r^3  \left\langle[T_\mu ^\nu (x),T_\mu^\nu(0)]\right\rangle e^{i \omega t}, 
\end{eqnarray}
where $\omega$ is the frequency of quark and gluon vibration. For a narrow frequency region, $\omega \rightarrow \omega _0 \equiv \omega_0 (T) \sim T$. 
From the relaxation time approximation, the deviation of energy-momentum tensor from its local equilibrium which is corresponding to the difference between the distribution function near and at equilibrium, $\delta n=n-n_0$, can be used to determine bulk and shear viscosities,  \cite{Tawfik:2010mb}
\begin{eqnarray}
\zeta (T,\mu, eB) &=&  \frac{3}{2\, T}   \sum_f  \frac{|q_f|B}{2\pi} \sum_\nu \int \frac{dp_z}{2\pi}  \left(2-\delta_{0\nu}\right) \,  \frac{\tau_f}{E_{B,f}^2} \left[\frac{|\vec{p}|^2}{3} - c_s^2 E_{B,f}^2 \right]^2 \, f_{f} (T, \mu) \left[1 + f_{f} (T, \mu) \right], \hspace*{5mm} \label{eq:gkbulk}\\
\eta (T,\mu, eB) &=& \frac{1}{5\, T} \,  \sum_f  \frac{|q_f|B}{2\pi} \sum_\nu \int \frac{dp_z}{2\pi}  \left(2-\delta_{0\nu}\right) \,  \frac{|\vec{p}|^4 \tau_f}{E_{B,f}^2} \; f_{f} (T, \mu) \left[1 + f_{f} (T, \mu)  \right]. \label{eq:gkshear}
\end{eqnarray}

The relaxation time ($\tau$) can be related to the quark decay constant ($\Gamma$). In doing this, two models can be implemented:
\begin{itemize}
\item The dynamic quasi-particle model (DQPM) which describes well the phenomenology of interacting massless quarks and gluons as non-interacting massive quasi-particles \cite{Peshier}. The flavor-blind reaction rates for quarks and gluons are inversely dependent on the particle width, $\tau_q \sim 1/\Gamma_q$ and $\tau_g \sim 1/\Gamma_g$,  respectively. 

\item The transport theory with Boltzmann master equation which can be used to calculate the momentum loss and the relaxation time, as well \cite{Zhuang95}. To this end, screening of long-range quark-quark interactions is assumed for the process of interpenetrating quark plasmas. For a spatially uniform quark plasma flowing with respect to $N_f-1$ plasmas with another type of flavors, the relative flow velocity is assumed to relax in time due to the collisions \cite{Marty13}.  In partonic and hadronic phases, the relaxation time, respectively, scales with the temperature $\tau \sim \exp [-T]$ for $T>T_c$  and $\tau\sim  1/T$ for $T<T_c$.
\end{itemize}

\begin{figure}[htb]
\centering{
\includegraphics[width=5cm,angle=-90]{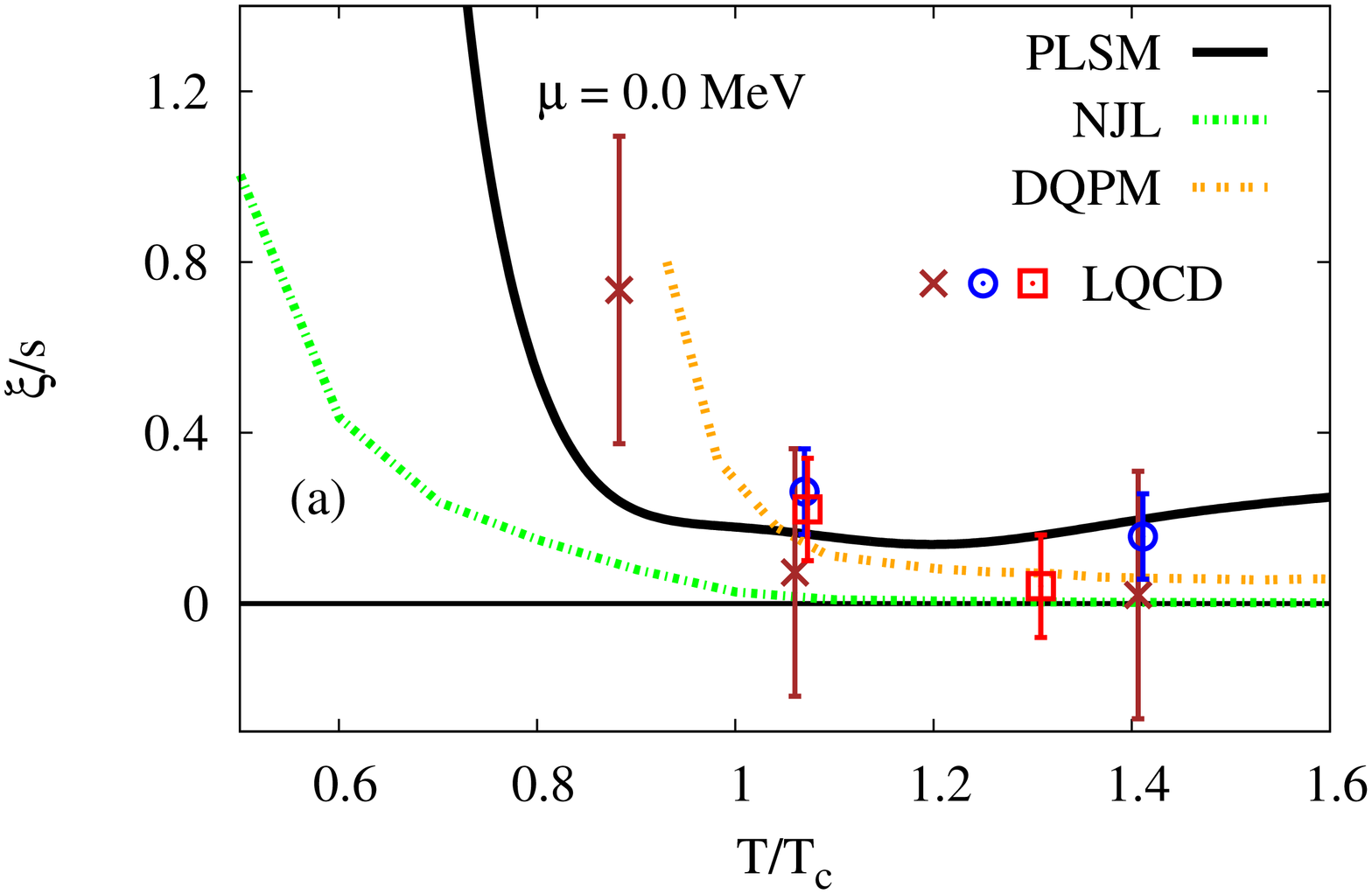}
\includegraphics[width=5cm,angle=-90]{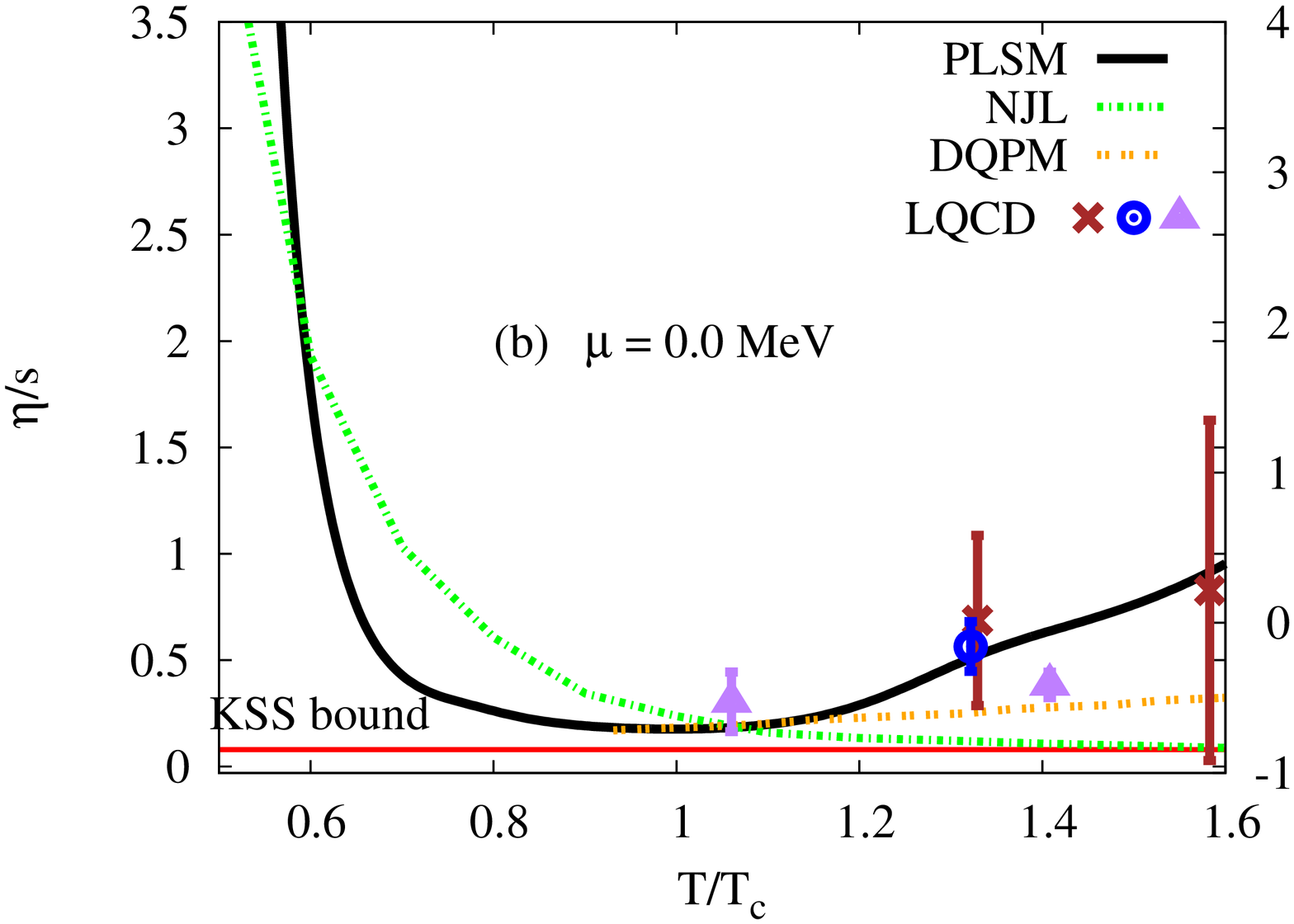} 
\caption{\footnotesize Left-hand panel: the ratio of bulk viscosity and thermal entropy ($\xi/s$) calculated from PLSM (solid curve) and compared with lattice QCD simulations  \cite{Meyer:2007} (cross points) and \cite{Sakai:2007} (circle points) and (square points) is given as function of temperature at vanishing chemical potential and magnetic field. Right-hand panel shows the ratio of shear viscosity to thermal entropy ($\eta/s$) calculated from PLSM (solid curve) and compared with the available lattice QCD  simulations  \cite{Meyer:2007} (cross points), \cite{Sakai:2007} (circle points) and (square points) and \cite{Sakai:2005} (triangle points). The Kovtun-Son-Starinets (KSS) bound is also depicted. Both panels are compared with NJL (dotted dash) and DQPM (double dotted) \cite{Bratkovskaya}. 
\label{viscosityB0}}}
\end{figure}

In vanishing magnetic field and at zero chemical potential, the temperature dependence of bulk viscosity ($\xi$) normalized to thermal entropy ($s$) is presented in left-hand panel of Fig. \ref{viscosityB0}. The ratio of shear viscosity to thermal entropy ($\eta/s$) is given in right-hand panel. At temperatures $\gtrsim T_c$, $\xi/s$ has a good agreement with the lattice calculations \cite{Meyer:2007,Sakai:2007,Sakai:2005,Karsch:2008}.  LSM \cite{Mao:2010}, DQPM and NJL  models \cite{Bratkovskaya} are also good with the lattice. The sharp raise of bulk viscosity slightly above the phase transition (as the entropy is assumed to remain constant) indicates instability in the hydrodynamic flow of the plasma as confirmed in RHIC observables \cite{Torrieri:2008}. The agreement between PLSM and lattice QCD calculations for $\eta/s$ is excellent above $T_c$.

\begin{figure}[htb]
\centering{
\includegraphics[width=5cm,angle=-90]{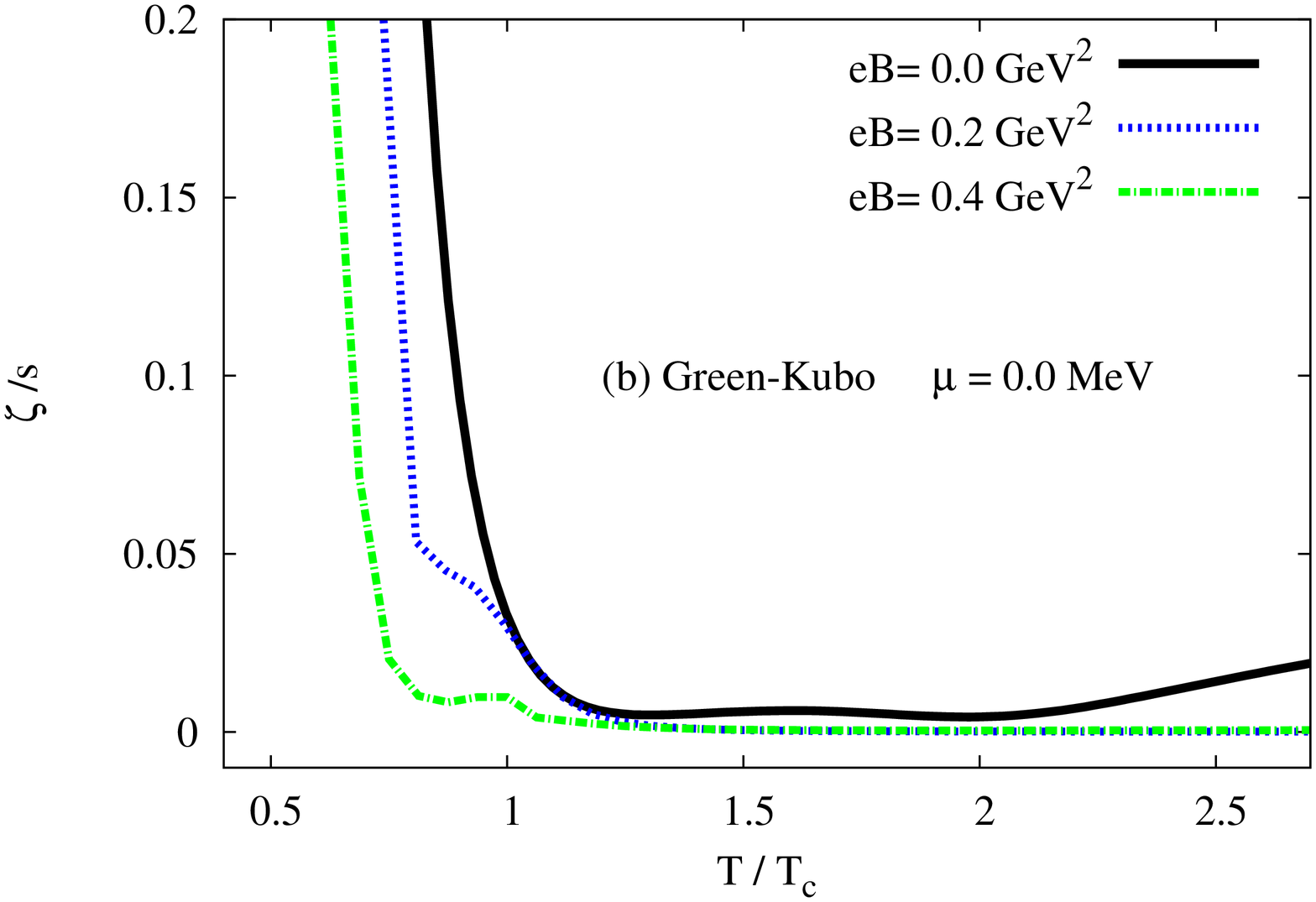}
\includegraphics[width=5cm,angle=-90]{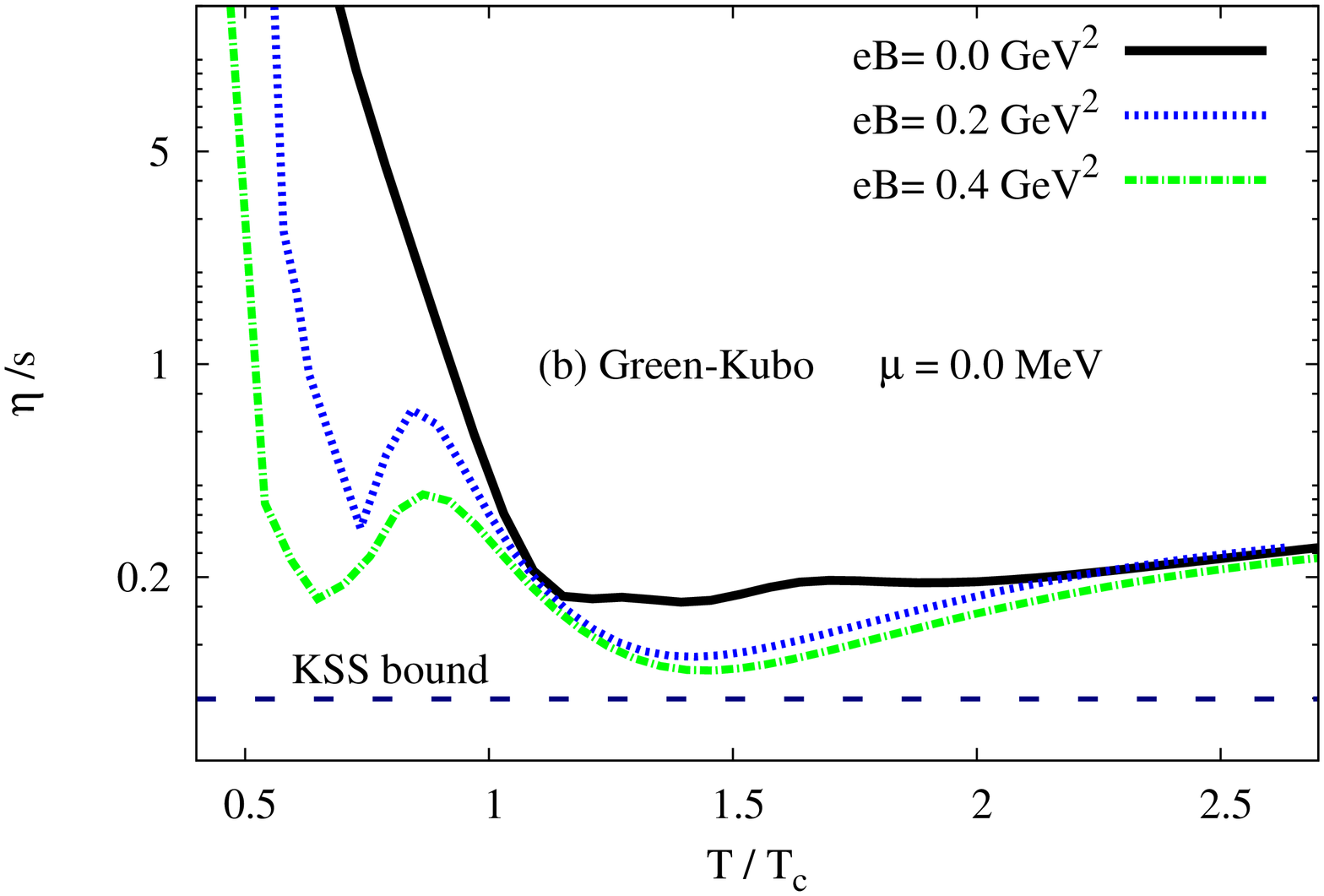} 
\caption{\footnotesize Left-hand panel: the influence of magnetic fields, $eB=0.0~$GeV$^2$ (solid), $eB=0.2~$GeV$^2$ (dotted) and $eB=0.4~$GeV$^2$ (dot-dashed)  on $\zeta/s$ is given as function of temperature at vanishing chemical potential. In right-hand panel, the same as in left-hand panel but for $\eta/s$. A very small increase can be observed at high temperature. \label{viscosityB}
}}
\end{figure}

The left-hand panel of Fig. \ref{viscosityB} shows the results of $\zeta/s$ vs. temperature in finite magnetic field. Again the results at $eB=0~$GeV$^2$ are depicted as reference (solid curve).  The results at $eB=0.0~$GeV$^2$ (solid), $eB=0.2~$GeV$^2$ (dotted) and $eB=0.4~$GeV$^2$ (dot-dashed) show a remarkable dependence on temperature. For instance, the finite magnetic field seems to enhance the appearance of the characterizing peak at $T_c$. The peaks are related to minima. At very low temperature, all curves seem to bundle together to large $\zeta/s$-values.

In right-hand panel of Fig. \ref{viscosityB}, the ratio $\eta/s$ is depicted as function of temperature at different values of $eB$. The solid curve represents the results at a vanishing magnetic field, while $eB=0.2$, $eB=0.15~$ and $0.4~$GeV$^2$ are given as dotted, dot-dashed  curves, receptively. KSS limit is represented by dashed line. At low temperatures, $\eta/s$ becomes very large. A peak positioned at $T_c$ starts to appear when the magnetic field increases while the peak's height is not sensitive to $e B$. The remarkable effect around $T_c$ refers to instability in the hydrodynamic flow of quarks and hadrons. At extremely high temperatures, the coupling becomes weak and the hadrons are entirely liberated into quarks and gluons. Thus, it is likely that the deconfinement phase-transition is completed. At temperatures $\gtrsim T_c$ the values of $\eta/s$ are not sensitive to the change in $e\, B$.

\begin{figure}[htb]
\centering{
\includegraphics[width=5cm,angle=-90]{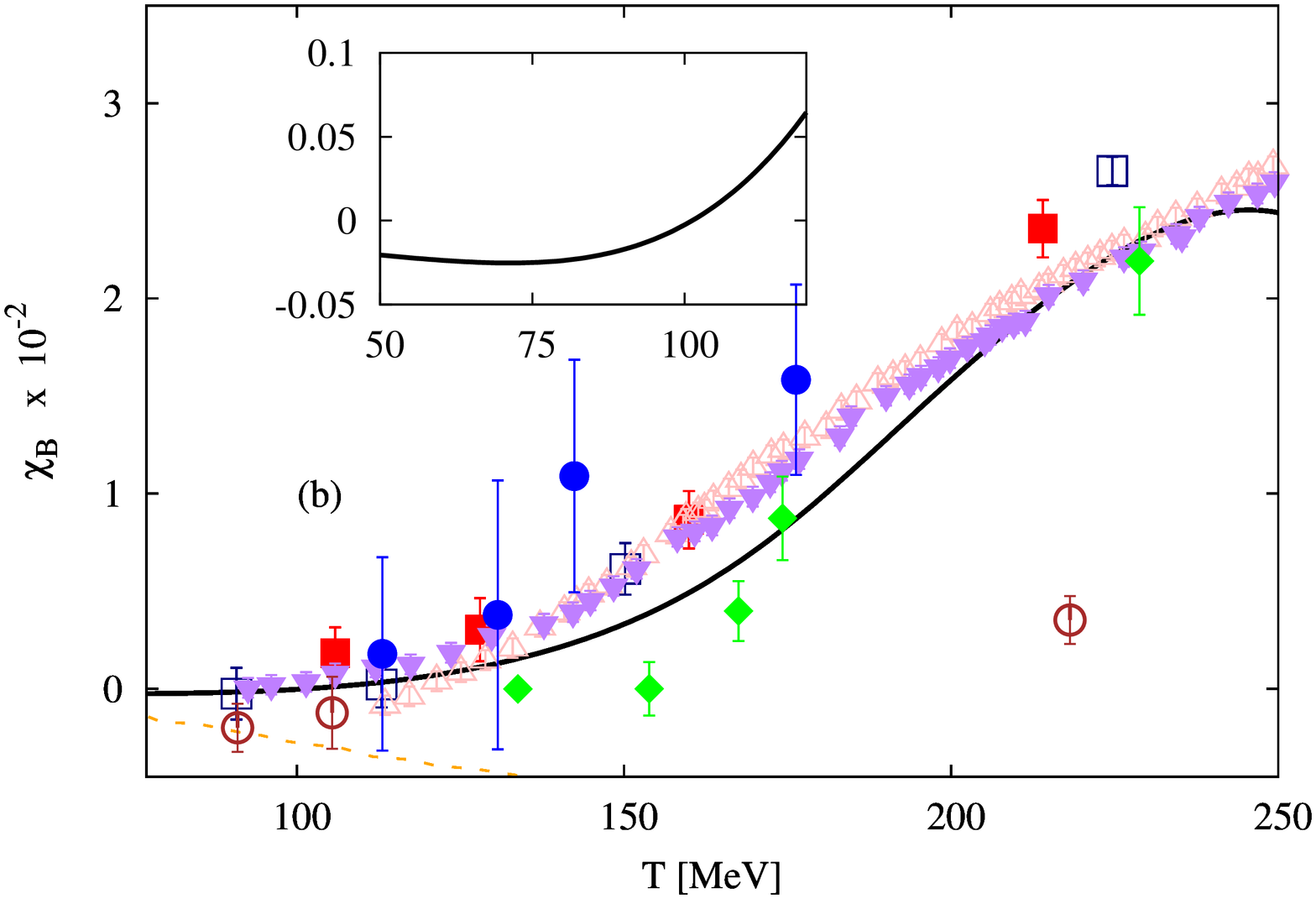}
\includegraphics[width=5.5cm,angle=-90]{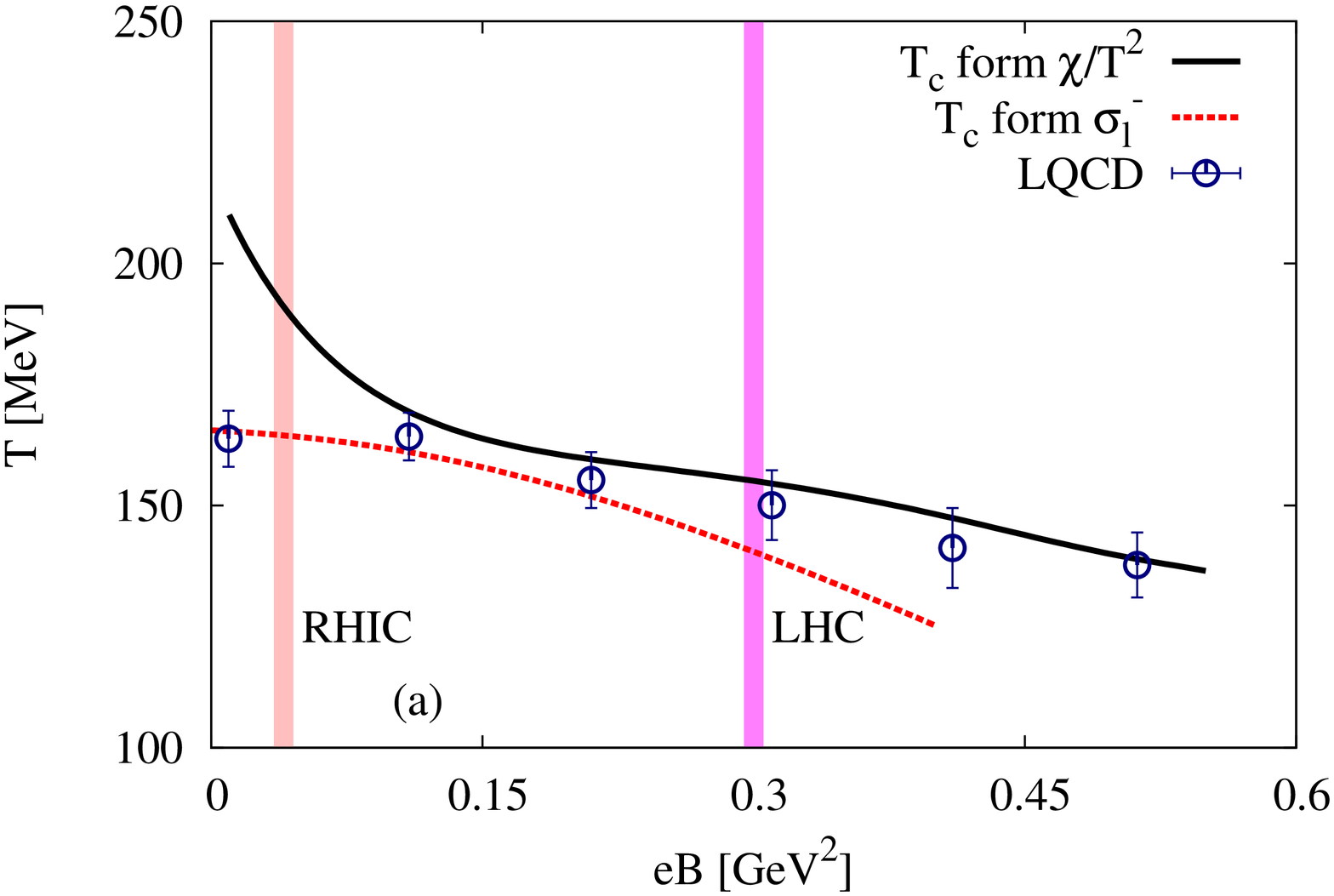}
\caption{\footnotesize Left-hand panel: the temperature dependence of magnetic susceptibility $\chi_B$ is given at $eB=0.2~$GeV$^2$ and vanishing baryon chemical potential. The results are compared with different lattice simulations (symbols) \cite{lattice:2014}.
Right-hand panel: the chiral phase-diagram relates the critical temperature to $eB$ through $\chi/T^2$ (solid curve) and quark condensate $\sigma_l$ (dashed curve). The vertical bands refer to magnetic fields at RHIC and LHC energies. 
\label{fig:eBdepence} 
}}
\end{figure}

The left-hand panel of Fig. \ref{fig:eBdepence} shows the magnetic susceptibility ($\chi_B$), which is determined from the second derivative of PLSM free energy with respect to $e B$, as function of temperature at $eB=0.2~$GeV$^2$ and  a vanishing baryon chemical potential. The magnetic susceptibility is a dimensionless quantity referring to the degree of magnetization as a response to the non-zero magnetic field. The results from PLSM are compared with various lattice simulations (symbols)  and with the HRG model. It worthwhile to notice the negative slope of $\chi_B$  (inside-box in middle panel). This signals that the QCD matter is diamagnetic. At high temperatures, i.e. restoring broken chiral symmetry, we observe a transition between dia- and para-magnetism. It is likely that the QCD matter is para-magnetic is very high temperatures. 

In right-hand panel, influences of the magnetic field strength on the chiral phase-diagram are analysed and compared with recent lattice QCD results (circles with errorbars) \cite{lattice:2014}. The vertical bands indicate the value of the magnetic fields at RHIC  ($\sim m_\pi ^2$) and LHC energies ($\sim 10-15\, m_\pi ^2$). There is a small suppression in the chiral condensates, i.e. magnetic catalysis. In PLSM we implement two methods in order to determine the quasi-critical temperature (dotted curve). The first one implements the higher-order moments of quark multiplicity, i.e. normalized quark susceptibility $\chi_q/T^2$ (solid curve). The second one uses the intersection of light-quark chiral condensate $\sigma_l$with the deconfinement order-parameters (dashed curve). We find that the critical temperature decreases with increasing $e B$, i.e. inverse magnetic catalysis. An excellent agreement is apparently achieved between our calculations and the lattice QCD, especially at small magnetic field $0\leq\,eB\,\leq 0.2$ GeV$^2$ (second method). The first method agrees well with the lattice data at a wider range of magnetic fields $0.13\leq\,eB\,\leq 0.55$ GeV$^2$. Again, the first method apparently overestimates lattice data at low temperature, while the second method slightly underestimates these at high temperature.

\section{Conclusion}

The magnetic field which is likely generated in peripheral heavy-ion collisions due to local imbalance in momentum carried out by the participants and off-center relativistic motion of spectator's electric charges seems to influence the transport properties (bulk and shear viscosity) in both hadron and quarks phases. Also, the quark-hadron phase transitions are remarkable influenced by such large magnetic field. 

In vanishing magnetic field, we find that increasing $T$ decreases $\zeta/s$. But at high temperatures, a small increase can be observed. In finite magnetic filed, there is a remarkable dependence on temperature; appearance of characterizing peaks at $T_c$ (enhanced with increasing $e B$) and very large values of $\zeta/s$ observed at low temperatures. For $\eta/s$, the peaks around $T_c$ gradually appear with increasing $e B$ but their heights seem not being sensitive to $e B$. At temperatures $\gtrsim T_c$, $\eta/s$ is also not sensitive to the change in $e\, B$.

The magnetic susceptibility $\chi_B$ which measures the degree of magnetization as a response of finite magnetic field seems to raise with the temperature. Both PLSM and lattice QCD calculations are in good agreement with each other. It is worthwhile to highlight the slightly negative $\chi_B$ takes place at low temperature. This refers to  diamagnetic properties of the QCD matter. Switching to high temperatures, i.e. restoring broken chiral symmetry, a transition between dia- and para-magnetic properties is observed.  At very high temperatures, the QCD matter is likely para-magnetic.
 
In nonzero magnetic field,  we have implemented two methods in order to determine the critical temperatures with changing $e B$. Magnetic catalysis is clearly observed in QCD matter. The agreement with the lattice calculations is fairly good.

\section*{Acknowledgements}
This work is partly supported by the World Laboratory for Cosmology And Particle Physics (WLCAPP), http://wlcapp.net/


\end{document}